\documentstyle[12pt,epsfig]{article}
\def\be{\begin{equation}} \def\ee{\end{equation}} \def\bea{\begin{eqnarray}}
\def\eea{\end{eqnarray}} \def\nnb{\nonumber}

\begin{document}
\renewcommand{\thefootnote}{\fnsymbol{footnote}}
\setcounter{footnote}{1}
 
\hfill{USC(NT)$-$Report$-$01$-$06}

\hfill{October 4, 2001}
\begin{center}
\vskip 1.0cm {\Large\bf
Analysis of ordinary and radiative muon capture \\
in liquid hydrogen }
\vskip 1.0cm
{\large Shung-ichi Ando\footnote{
E-mail address : sando@nuc003.psc.sc.edu},
Fred Myhrer\footnote{
E-mail address : myhrer@sc.edu},
and Kuniharu Kubodera\footnote{
E-mail address : kubodera@sc.edu}
}\\
\vskip 0.8cm
{\large {\it Department of Physics and  Astronomy, 
University of South Carolina,\\
Columbia, SC 29208, USA}} \\
\end{center}
\vskip 1.0cm
A simultaneous analysis is made
of the measured rates of ordinary muon capture (OMC) 
and radiative muon capture (RMC) in liquid hydrogen,
using theoretical estimates for 
the relevant atomic capture rates 
that have been obtained in chiral perturbation theory
with the use of the most recent values 
of the coupling constants.
We reexamine the basic formulas 
for relating the atomic OMC and RMC rates 
to the liquid-hydrogen OMC and RMC rates, respectively.
Although the analysis is significantly influenced
by ambiguity in the molecular state population,
we can demonstrate that, 
while the OMC data can be reproduced,
the RMC data can be explained only with unrealistic values of the 
coupling constants;
the degree of difficulty becomes even more severe
when we try to explain the OMC and RMC data
simultaneously.


\renewcommand{\thefootnote}{\arabic{footnote}}
\setcounter{footnote}{0}
\newpage
\noindent
{\bf 1. Introduction}

Ordinary and radiative muon capture (OMC and RMC) on a proton 
\bea
\mu^- + p \to n+\nu_\mu, \ \ \ \ \ \ {\rm (OMC)}\\
\mu^- + p \to n+\nu_\mu +\gamma ,  \ \ \ {\rm (RMC)}
\eea
are fundamental weak-interaction processes 
in nuclear physics 
and a primary source of information on $g_P$, 
the induced pseudoscalar coupling constant 
of the weak nucleon current, 
see {\it e.g.} \cite{morita,bernard-axial-review}.
The most accurate existing measurements 
of the OMC and RMC rates have been carried out
using a liquid hydrogen target, 
which unfortunately makes  
the analysis of the data
sensitive to the molecular transition rates
in liquid hydrogen.
We denote by $\Lambda_{liq}$ the OMC rate
in liquid hydrogen.
The experimental value obtained by 
Bardin {\it et al.}~\cite{omc-exp} is
\bea
\Lambda_{liq}^{exp} = 460 \pm 20 
\ \ \ \mbox{\rm [s$^{-1}$]}   
\ \ \ \ \ \ {\rm (OMC)} .
\label{eq;exp-omc}
\eea
As for RMC, Jonkmans {\it et al.}~\cite{rmc-exp} 
measured the absolute photon spectrum for 
$E_\gamma\ge 60$ MeV and deduced therefrom
the partial RMC branching ratio, $R_\gamma$, 
which is the number of RMC events (per stopped muon)
producing a photon with $E_\gamma\ge 60$ MeV.
The measured value of $R_\gamma$ is \cite{rmc-exp,wright}
\bea
R^{exp}_\gamma = (2.10\pm 0.22)\times 10^{-8}
\ \ \ {\rm (RMC)}  .
\label{eq;exp-rmc}
\eea
Surprisingly, the value of $g_P$
deduced in \cite{rmc-exp,wright}
from the RMC data
is $\sim$1.5 times larger than the PCAC prediction~\cite{PCAC}.
By contrast, the value of $g_P$ deduced in \cite{bardin}
from the OMC data is in good agreement with 
the PCAC prediction. 

On the theoretical side, 
the early estimation of $g_P$ was made using PCAC.
Heavy-baryon chiral perturbation theory (HB$\chi$PT),
a low-energy effective theory of QCD,
allows us to go beyond the PCAC approach,
but the results of detailed HB$\chi$PT 
calculations~\cite{bernard-gp} 
up to next-to-next-to-leading order (NNLO)
essentially agree with those obtained in the PCAC approach.
Thus the theoretical framework for 
estimating $g_P$ is robust.
The key quantities in analyzing OMC and RMC
are the atomic rates, $\Lambda_s$ and $\Lambda_t$,
where $\Lambda_s$ ($\Lambda_t$) is the capture rate
for the hyperfine singlet (triplet) state 
of the $\mu$-$p$ atom.
\footnote{The rates $\Lambda_s$ and $\Lambda_t$
are generic symbols 
for OMC and RMC. 
When we need to distinguish
OMC and RMC, we use the symbols 
$\Lambda_{s,t}^{OMC}$ 
and $\Lambda_{s,t}^{RMC}$.} 
The atomic rates for OMC and RMC have also been estimated 
in the framework of 
HB$\chi$PT~\cite{fearing-omc,AMK,bernard-fearing98,fearing98,
meissner,AM1,bernard-omc-rmc}.
The expressions obtained in HB$\chi$PT
have been found to be essentially in agreement with 
those of the earlier 
work~\cite{opat,primakoff,fearing80,gumitro81,bf87}.
It has also been confirmed that the chiral expansion 
converges rapidly, rendering 
estimates of the OMC and RMC rates obtained in $\chi$PT 
extremely robust.
As for the numerical results, however,
the earlier estimates of the atomic OMC rates, {\it e.g.}  
\cite{opat,primakoff}, need to be revised
because some values of the input parameters 
($g_A$, $g_{\pi N}$, {\it etc.}) used in those estimates
are now obsolete.
In Ref.~\cite{AMK}, we provided updated estimates of 
$\Lambda_s^{OMC}$ and $\Lambda_t^{OMC}$ 
based on HB$\chi$PT (up to NNLO). 
A notable finding in \cite{AMK} is 
that the use of the recent larger value of 
the Gamow-Teller coupling constant, $g_A$,
gives a value of $\Lambda_s^{OMC}$ that is significantly larger 
than the older value commonly quoted
in the literature, see Refs.~\cite{AMK,bernard-omc-rmc}.

To make comparison between theory and experiment,
one needs to relate the theoretically calculated 
atomic OMC and RMC rates 
to $\Lambda_{liq}$ and $R_\gamma$, respectively.
For convenience, we refer to this relation
as the A-L (atom-liquid) formula. 
Bakalov {\it et al.}~\cite{bakalov} 
made a detailed study of
the A-L formula,  
and they gave an explicit expression for $\Lambda_{liq}$ 
(see Eq.~(56c) in Ref.~\cite{bakalov}). 
In our previous work \cite{AMK} we analyzed $\Lambda_{liq}$
using the A-L formula of Bakalov {\it et al.}
and found that the best available estimates 
of the atomic capture rates based on HB$\chi$PT
would lead to a value of $\Lambda_{liq}$ that was 
significantly larger than $\Lambda_{liq}^{exp}$.
We also reported that, by introducing 
a molecular state mixing parameter, $\xi$, 
considered by Weinberg~\cite{wei},
it was possible to reproduce $\Lambda_{liq}^{exp}$ 
and $R_\gamma^{exp}$ simultaneously. 
However, the A-L formula of Bakalov {\it et al.} 
does not correspond to the experimental condition of OMC;
to compare with $\Lambda_{liq}^{exp}$,  
the time sequence
of the experimental measurement 
should be considered \cite{omc-exp}.

In this work we reexamine 
$\Lambda_{liq}$ and $R_\gamma$
by incorporating into our analysis
the experimental conditions 
as well as the updated estimates 
of the atomic capture rates.
In particular, we investigate 
the influence of ambiguity in the transition rate 
between the molecular states (to be discussed below);
we also examine the dependence of the results
on the values of $g_P$ and 
the molecular parameter $\xi$ (see below). 

\vskip 2mm \noindent
{\bf 2. Muonic states in liquid hydrogen}

To evaluate $\Lambda_{liq}$ and $R_\gamma$ 
from the calculated atomic OMC and RMC rates,
we need to know the temporal behavior of
the various $\mu$-capture components
(capture from the atomic states
and capture from $p$-$\mu$-$p$ molecular states).
\begin{figure}
\begin{center}
\epsfig{file=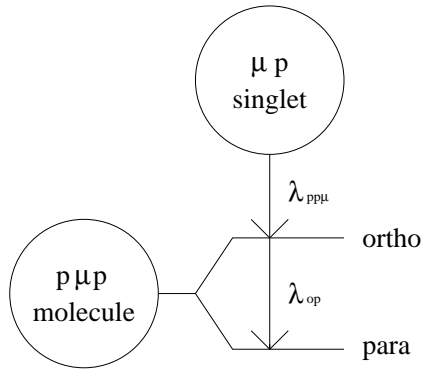,width=5.5cm}
\caption{
Atomic and molecular states relevant to 
muon capture in liquid hydrogen;
$\lambda_{pp\mu}$ is the transition rate 
from the atomic singlet state
to the ortho $p$-$\mu$-$p$ molecular state, 
and $\lambda_{op}$ is that
from the ortho to para molecular state.}
\label{fig;liquid}
\end{center}
\end{figure}
Fig.~\ref{fig;liquid} 
schematically depicts various competing atomic 
and molecular processes occurring in liquid hydrogen. 
A muon stopped in liquid hydrogen quickly forms 
a muonic atom ($\mu$-$p$) in the lowest Bohr state.
The atomic hyperfine-triplet state (S=1) 
decays extremely rapidly
to the singlet state (S=0),  
with a transition rate
$\lambda_{10}\simeq 1.7 \times 10^{10}$ s$^{-1}$.
In the liquid hydrogen target
a muonic atom and a hydrogen molecule 
collide with each other
and form a $p$-$\mu$-$p$ molecule 
with the molecule predominantly in its ortho state.
We denote by $\lambda_{pp\mu}$  
the transition rate from the atomic singlet state 
to the ortho $p$-$\mu$-$p$ molecular state.
The ortho $p$-$\mu$-$p$ state further decays to the para 
$p$-$\mu$-$p$ molecular state. 
This rate is denoted by $\lambda_{op}$. 
Let $N_s(t)$, $N_{om}(t)$, and $N_{pm}(t)$
represent the numbers of muons at time $t$ 
in the atomic singlet, ortho-molecular, 
and para-molecular states, respectively.
They satisfy coupled kinetic equations,
see Eq.~(54a) in Ref.~\cite{bakalov}. 
To integrate these coupled differential equations,
we need to know the initial conditions. 

For illustration purposes, let us consider a case in which
there is one muon in the singlet state at $t=0$;
{\it i.e.}, $N_s(0)=1$ and $N_{om}(0)=N_{pm}(0)=0$.
We then have
\bea
N_s(t) &=& e^{-\lambda_2 t} ,
\ \ \
N_{om}(t) = \frac{\lambda_{pp\mu}}{\lambda_2-\lambda_3}
(e^{-\lambda_3t}-e^{-\lambda_2t}) ,
\nnb \\
N_{pm}(t) &=& 
\frac{\lambda_{op}\lambda_{pp\mu}}
{(\lambda_3-\lambda_4)(\lambda_2-\lambda_4)}
e^{-\lambda_4t} 
-\frac{\lambda_{op}\lambda_{pp\mu}}
{(\lambda_2-\lambda_3)(\lambda_3-\lambda_4)}
e^{-\lambda_3t} 
\nnb \\ &&
+\frac{\lambda_{op}\lambda_{pp\mu}}
{(\lambda_2-\lambda_3)(\lambda_2-\lambda_4)}
e^{-\lambda_2t} ,
\label{eq;muons}
\eea
where 
$\lambda_2 = \lambda_0 + \lambda_{pp\mu}
+\Lambda_s^{OMC}+\Lambda_s^{RMC}$,
$\lambda_3 = \lambda_0 + \lambda_{op}
+\Lambda_{om}^{OMC}+\Lambda_{om}^{RMC}$, 
$\lambda_4 = \lambda_0 + \Lambda_{pm}^{OMC}
+\Lambda_{pm}^{RMC}$.
Here $\lambda_0$ is the muon natural decay rate.
$\Lambda_{om}^{OMC}$ and $\Lambda_{pm}^{OMC}$ are
the OMC rates in the ortho molecular 
and para molecular states, respectively;
similarly for $\Lambda_{om}^{RMC}$, 
and $\Lambda_{pm}^{RMC}$.\footnote{Since 
$\Lambda^{RMC}_{s}$, $\Lambda^{RMC}_{om}$ $\Lambda^{RMC}_{pm}$ 
are very small, 
they can be ignored in the calculation 
of $N_{s,om,pm}(t)$.
In evaluating the RMC rate itself,
however, we need these capture rates;
see Eq.~(\ref{eq;Ngamma}) below.}
These rates are given by
\bea
\Lambda_{om}^F = 2\gamma_O
\left(\frac34\Lambda_s^F+\frac14\Lambda_t^F\right),
\ \ \ 
\Lambda_{pm}^F = 2\gamma_P
\left(\frac14\Lambda_s^F+\frac34\Lambda_t^F\right),
\label{eq:gamop}
\eea
where $F$ stands for ``$OMC$" or ``$RMC$",
and $2\gamma_O=1.009$, 
$2\gamma_P=1.143$ \cite{bakalov}.

At this point we discuss the numerical values of
$\lambda_{pp\mu}$ and $\lambda_{op}$. 
The former shows a wide scatter in the literature,
ranging from 
$\lambda_{pp\mu}=(1.89\pm 0.20)\times 10^6$ s$^{-1}$ 
to $(2.75\pm 0.25)\times 10^6$ s$^{-1}$ \cite{PSI}.
In this work, for the sake of definiteness,
we employ the averaged value 
$\lambda_{pp\mu} = 2.5 \times 10^6$ s$^{-1}$
(the main point of our argument is not affected
by this choice). This value is comparable to 
the muon decay rate 
$\lambda_0=0.455\times 10^6$ s$^{-1}$. 
As regards $\lambda_{op}$, 
there is a significant difference between
the experimental and theoretical values;
$\lambda_{op}^{exp}=
(4.1\pm1.4)\times 10^4$ s$^{-1}$ \cite{bardin} 
as compared with 
$\lambda_{op}^{th} = (7.1\pm 1.2) \times 10^4$ s$^{-1}$ 
\cite{bakalov}. 

The dominant state for the OMC and RMC measurements 
is the ortho molecular state as 
is evident from Eq.~(\ref{eq;muons}).
In both measurements, 
data taking starts at $t=t_i\ne 0$,
and it is essential to incorporate this aspect 
into the A-L formula (see below). 
Furthermore, in the OMC experiment 
the time dependence of the population
of each state plays an important role.

\vskip 2mm \noindent
{\bf 3. Atom-Liquid (A-L) formula for OMC and RMC}

The discussion so far is common 
for both OMC and RMC,
but we now turn to the individual discussion
of each case.
In the OMC experiment (see Fig.~4 in Ref.~\cite{omc-exp}),
$\mu^-$ beams arrive at the target on the average 
in a 3 $\mu$s-long burst with repetition rate 3000 Hz.
The data collection typically starts 
1 $\mu$s after the end of the 3 $\mu$s-long beam burst,
and the measurement lasts until
306 $\mu$s after the end of the beam burst.  
As mentioned, the cascade processes leading
to the $\mu$-$p$ ground state 
and the transition between the 
atomic hyperfine states are extremely fast. 
One therefore can safely ignore a time lag 
between the muon arrival time and the time
at which the $\mu$-$p$ atomic hyperfine-singlet 
state is formed. 
To proceed with the consideration of OMC, 
we assume 
that the average time intervals of Ref.~\cite{omc-exp} 
cited above are actual time intervals. 
Then, provided all the muons arrive at the same time,
we can choose with no ambiguity 
that arrival time as the origin of time ($t=0$)
and let $t$ = $t_i$, the starting time 
for data collection, refer to that origin. 
However, the finite duration ($t_b$ = 3 $\mu$s) 
of the beam burst causes uncertainty 
in the value of $t$ = $t_i$ 
to be used in Eq.~(\ref{eq;muons});   
$t_i$ can be anywhere between 1.0 $\mu$s and 4.0 $\mu$s. 
To account for this muon pulse duration time, $t_b$, 
we assume for simplicity 
that the beam pulse has a rectangular shape. 
Then, at time $t$ the average number of residual muons are:  
\bea
\bar{N}_\mu(t) &\equiv& 
\frac{1}{t_b}\int^{t_b}_0dt^\prime
N_\mu(t-t^\prime).
\eea 
where $N_\mu(t)=N_s(t)+N_{om}(t)+N_{pm}(t)$. 
The OMC experiment \cite{omc-exp} 
counts the number of electrons
produced by $\mu^- \to e^-\bar{\nu}_e\nu_\mu$, 
and $\Lambda_{liq}$
is deduced from the difference between the muon decay rate in 
liquid hydrogen and that in vacuum;
the latter is determined from the number of positrons produced in 
$\mu^+ \to e^+ \nu_e \bar{\nu_\mu}$.
We use the expression of 
Ref.~\cite{omc-exp} (and $t_i=4$ $\mu$s) 
\bea
\Lambda_{liq} &\equiv& \left(
\frac{\int^\infty_{t_i}dt \frac{d\bar{N}_e}{dt}}
 {\int^\infty_{t_i}dt(t-t_i)
 \frac{d\bar{N}_e}{dt}}\right)-\lambda_0 ,
\label{eq;AL-OMC}
\eea
where 
$\bar{N}_e(t)$ is the averaged number of 
electrons produced at time $t$ and 
$\frac{d\bar{N}_e(t)}{dt} = \lambda_0 \bar{N}_\mu(t)$. 
Here we have used the fact the duration 
of the measuring time (306 $\mu$s)
is long enough to be treated as $\infty$.

On the other hand, for the RMC experiment 
\cite{rmc-exp,wright},
the muons essentially arrive one by one 
and 
the data taking begins at $t_i=365$ ns.
We therefore can neglect the beam burst duration time 
in the RMC case, and we obtain 
\bea
R_\gamma &=& 
\frac{N_\gamma(\infty)-N_\gamma(t_i)}
 {N_\mu(t_i)}  . 
\label{eq;AL-RMC}
\eea
Here $N_\gamma(t)$ is the number of photons 
obtained by integrating the photon spectrum 
over the interval, $60\le E_\gamma \le 99$ MeV, and 
the production of photons in RMC is determined by 
\bea 
\frac{d N_\gamma (t)}{dt} = 
\Lambda_s^{RMC} N_s(t) + 
\Lambda_{om}^{RMC} N_{om}(t) 
+\Lambda_{pm}^{RMC} N_{pm}(t)\,,
\label{eq;Ngamma}  
\eea 
where $N_\gamma(0) = 0$.

\vskip 2mm \noindent
{\bf 4. Numerical results and discussion}
\begin{table}
\begin{center}
\begin{tabular}{|cc||cccc|} \hline
$g_A$ & $g_{\pi N}$ & 
$\Lambda_s^{OMC}$ & $\Lambda_t^{OMC}$ &
$\Lambda_s^{RMC}$ & $\Lambda_t^{RMC}$ \\ \hline
1.267 & 13.40 &
695 & 11.9 & $0.891\times 10^{-3}$ & $20.1\times 10^{-3}$ \\ \hline
\end{tabular}
\caption{Coupling constants and the atomic capture rates [s$^{-1}$] 
used in the present analysis 
}
\label{table;rates}
\end{center}
\end{table}

We first give the numerical values of inputs
to be used in what follows.
Table \ref{table;rates} presents
the values of the coupling constants
and the atomic capture rates.
The OMC and RMC rates 
for the hyperfine-singlet and -triplet states 
have been calculated in HB$\chi$PT 
up to NNLO~\cite{AMK,AM1} and 
with the use of the most recent values 
of the coupling constants discussed in \cite{AMK}.

We estimate $\Lambda_{liq}$ 
by using the atomic OMC rates 
in Table \ref{table;rates}.
Besides the A-L formula in Eq.~(\ref{eq;AL-OMC}),
we consider two others for the sake of comparison; 
these two A-L formulas are 
that of Bardin {\it et al.}~\cite{bardin} 
and that of Bakalov {\it et al.}~\cite{bakalov}. 
For the ortho-para transition rate
we employ either $\lambda_{op}^{exp}$ or $\lambda_{op}^{th}$. 
The use of $\lambda_{op}^{exp}$
leads to $\Lambda_{liq}$ = 460 s$^{-1}$ 
with Eq.~(\ref{eq;AL-OMC}) 
and $\Lambda_{liq}=459$ s$^{-1}$ with Bardin {\it et al.}'s A-L formula.
Those values agree with 
$\Lambda_{liq}^{exp}=460\pm 20$ s$^{-1}$.
Meanwhile, if we employ $\lambda_{op}^{th}$,
we obtain 
$\Lambda_{liq}$ $\simeq$ 421 s$^{-1}$
with Eq.~(\ref{eq;AL-OMC}),
and
$\Lambda_{liq}$ $\simeq$ 419 s$^{-1}$
with Bardin {\it et al.}'s formula. 
Thus, $\Lambda_{liq}$ is highly sensitive
to $\lambda_{op}$.  
On the other hand, the use of Bakalov {\it et al.}'s
A-L formula \cite{bakalov} 
gives too large a value for $\Lambda_{liq}$
regardless of whether we use 
$\lambda_{op}^{exp}$ or $\lambda_{op}^{th}$\,;
$\Lambda_{liq}$=532 s$^{-1}$ for $\lambda_{op}^{exp}$,
and 
$\Lambda_{liq}$=518 s$^{-1}$ for $\lambda_{op}^{th}$.
As mentioned before, the A-L formula of Bakalov {\it et al.}, 
which was adopted in our previous work~\cite{AMK}, 
corresponds to the choice of $t_i=0$,
and this choice does not simulate the experimental condition.

An estimate of $R_\gamma$ is obtained 
from Eq.~(\ref{eq;AL-RMC})
and the atomic RMC rates 
in Table \ref{table;rates}.
With the use of $\lambda_{op}^{exp}$, 
the calculated value of $R_\gamma$ is significantly smaller 
than $R_\gamma^{exp}$ in Eq.~(\ref{eq;exp-rmc});
$R_\gamma^{exp}/R_\gamma^{th}\approx 1.5$.
If in Eq.~(\ref{eq;AL-RMC}) 
we use $\lambda_{op}^{th}$ instead of 
$\lambda_{op}^{exp}$, then $R_\gamma$ is enhanced 
by about 9\% but the increase 
is not large enough to reconcile
$R_\gamma^{th}$ with $R_\gamma^{exp}$. 
Thus it is not possible to reproduce
$R_\gamma^{exp}$ in the existing theoretical framework
with the use of the standard set of input parameters.
In addition, we remark 
that our results indicate
that the sensitivity of $R_\gamma$ to $\lambda_{op}$
is comparable to that of $\Lambda_{liq}$.

As mentioned, the atomic capture rates
calculated using a phenomenological 
relativistic tree-level model~\cite{fearing80}
are consistent with those of HB$\chi$PT
\cite{meissner,AM1,bernard-omc-rmc,AFM}
(provided the former uses the updated value of $g_A$ 
and the PCAC value of $g_P$).
Therefore, the above conclusions 
are not necessarily unique to HB$\chi$PT.
However, since HB$\chi$PT gives 
$\Lambda_s$ and $\Lambda_t$ with high precision
(primarily because the value of $g_P$ is 
strictly restricted by chiral symmetry),
it allows us to draw much sharper conclusions
than the phenomenological approach.

Next, we discuss the sensitivity of 
$\Lambda_{liq}$ and $R_\gamma$ to possible changes in the values
of $g_P$ and the molecular mixing parameter $\xi$. 
In this discussion we use the
phenomenological model of Fearing~\cite{fearing80},
a model which admits the variation of $g_P$
within a certain range. 
As discussed by Weinberg~\cite{wei},
the possible mixing of the ortho molecular 
$p$-$\mu$-$p$ spin 3/2 state and spin 1/2 state,
parameterized by $\xi$, 
may change the molecular capture rates to
\bea
\Lambda_{om}'^F= \xi 
\Lambda_{om}^F(1/2) + (1-\xi) \Lambda_{om}^F(3/2),
\eea
where $F$ stands for ``OMC" or ``RMC";
$\Lambda_{om}^F(1/2) = \Lambda_{om}^F$ 
[see Eq.~(\ref{eq:gamop})] and 
$\Lambda_{om}^F(3/2) = 2\gamma_O\Lambda_t^F$.
Although the existing theoretical estimate favors
$\xi\simeq 1$~\cite{bakalov,halpern},
we treat it here, as we did in Ref.~\cite{AMK},
as a parameter to fit the data.
In this phenomenological model, 
with the use of $\lambda_{op}^{exp}$, 
we can reproduce $R_\gamma^{exp}$ 
by adopting either $g_P = 1.4 g_P^{PCAC}$ or $\xi=0.80$. 
However, with the same value of $\lambda_{op}$, 
the OMC data requires  
$g_P \le 1.2 g_P^{PCAC}$ or $\xi \ge 0.95$.  
Therefore, it is impossible to simultaneously 
fit the OMC and the RMC data even 
by adjusting $g_P$ and $\xi$. 
If $\lambda_{op}$ is taken to be smaller than 
$\lambda_{op}^{exp}$ = 4.1$\times$10$^6$ s$^{-1}$, 
then it is not impossible to explain
$\Lambda^{exp}_{liq}$ and $R^{exp}_\gamma$ 
within the phenomenological model 
with a value of $g_P$ larger than that of $g_P^{PCAC}$
and $\xi\le 0.95$.
However, we cannot attach too much significance 
to this possibility,
since HB$\chi$PT constrains 
the value of $g_P$ with high accuracy,
{\it i.e.}, there is not much room left for adjusting 
the value of $g_P$.
The result of a more precise measurement 
of $\lambda_{op}$ at TRIUMF \cite{Lop-experiment}
will shed much light on this issue. 

Our findings are largely in the nature
of reconfirming the conclusions stated in one way 
or another in the literature, but a coherent
treatment of OMC and RMC in liquid hydrogen 
as described here is hoped to be useful.
Our treatment is characterized 
by the use of the best available atomic capture rates
obtained in HB$\chi$PT, and by an improved A-L formula.
Although we have presented examples
of simulation of the experimental conditions,
they are only meant to serve illustrative purposes.
Definitive analyses can be done only by the people
who carried out the relevant experiments.
Finally, we remark that
a precise measurement of the OMC rate in hydrogen gas
is planned at PSI \cite{PSI}. This experiment would 
eliminate the ambiguity of the molecular transition rate 
discussed in this paper and 
directly test the HB$\chi$PT prediction \cite{AMK,bernard-omc-rmc}. 

\vskip 2mm \noindent
{\bf Acknowledgment}

This work was motivated by Dr.~T. Gorringe's criticism
(communicated to us by Dr.~H.W. Fearing)
about the assumption $t_i=0$ made in our earlier work.
We are deeply grateful to these two colleagues
for that information and also for other illuminating 
remarks.
Thanks are also due to Dr.~T.-S. Park 
for useful discussions.
This work is supported in part by the U.S. 
National Science Foundation, Grant No. PHY-9900756
and Grant No. INT-9730847.

\end{document}